\documentclass[aps,prb,twocolumn,floatfix,showpacs]{revtex4-1}
\usepackage{graphicx}
\usepackage{amsmath}
\usepackage{bm}
\usepackage{verbatim}
\usepackage{color}
\usepackage{ulem}
\def\be{\begin{equation}}
\def\ee{\end{equation}}
\def\bea{\begin{eqnarray}}
\def\eea{\end{eqnarray}}
\def\kv{{\vec{k}}}

\def\l{\ell}
\def\tv{t_{v}}
\def\dg{\dagger}
\def\no{\nonumber}
\def\e{{\varepsilon}}

\def\a{{\alpha_\kv}}
\def\b{{\beta_\kv}}
\def\c{{\alpha_{m\kv}}}
\def\d{{\beta_{m\kv}}}

\def\Iv{{\bf I}}

\newcommand{\bra}[1]{\left\langle{#1}\right|}
\newcommand{\ket}[1]{\left|{#1}\right\rangle}
\newcommand{\braket}[1]{\langle{#1}\rangle}

\begin{document}

\title{Entanglement spectrum of fermionic bilayer honeycomb lattice:\\ Hofstadter butterfly}

\author{Z. Moradi}
\affiliation{Department of Physics, Institute for Advanced Studies
in Basic Sciences (IASBS), Zanjan 45137-66731, Iran}

\author{J. Abouie}
\email{jahan@iasbs.ac.ir}
\affiliation{Department of Physics, Institute for Advanced Studies
in Basic Sciences (IASBS), Zanjan 45137-66731, Iran}

\begin{abstract}
We perform an analytical study of the energy and entanglement spectrum of
non-interacting fermionic bilayer honeycomb lattices in the presence
of trigonal warping in the energy spectrum, on-site energy difference and uniform magnetic field.
Employing single particle correlation functions, we present an explicit form for layer-layer entanglement Hamiltonian whose
spectrum is entanglement spectrum. We demonstrate that in the
absence of trigonal warping, at zero on-site energy difference exact
correspondence is established between entanglement spectrum and
energy spectrum of monolayer which means that the entanglement
spectrum perfectly reflects the edge state properties of the
bilayer. We also show that trigonal warping breaks down such a perfect
correspondence, however, in $\Gamma$-K direction in hexagonal
Brillouin zone, their behaviors are remarkably the same for
particular relevances of hopping parameters. In the presence of an
on-site energy difference the symmetry of entanglement spectrum is
broken with opening an indirect entanglement gap. We also study the
effects of a perpendicular magnetic field on both energy and the entanglement
spectrum of the bilayer in the presence of trigonal warping and on-site energy difference.
We demonstrate that the entanglement spectrum versus magnetic flux has a self similar fractal structure, known Hofstadter butterfly.
Our results also show that the on-site energy
difference causes a transition from the Hofstadter butterfly to a
tree-like picture.
\end{abstract}

\date{\today}
 \maketitle

\section{Introduction}\label{sec:intro}
Entanglement is a kind of nonlocal correlations that plays key role
in condensed matter physics\cite{Amico08}. The study of entanglement
provides new insights into topological states of matter, which can
not be characterized using local order parameters. For instance, the
topological entanglement entropy of von Neumann entanglement
entropy,  the most commonly used measurement of entanglement, is
directly related to the total quantum dimension of fractional
quasiparticles\cite{Levin06, Kitaev06}. The full {\it entanglement
spectrum} (ES) provides even more complete information than
topological entanglement entropy\cite{Li08}. The entanglement
spectrum of a bipartite system with $A$ and $B$ subsystems, can be
characterized by the eigenvalues of the reduced density matrix
$\rho_{red.}$ of either one of the two subsystems in the ground
state of composite system. The reduced density matrix of a subsystem is obtained by tracing out the
other subsystem degrees of freedom. As
the eigenvalues of the reduced density matrix are non-negative, one
can write $\rho_{red.}=\frac 1Z e^{-{\cal H}}$,
 where $Z=tr(e^{-{\cal H}})$ is a partition function at temperature $T=1$ and ${\cal H}$ is the {\it entanglement Hamiltonian} whose spectrum is the ES.

The notion of entanglement spectrum has now been applied to many
different systems. These include quantum Hall monolayers at
fractional
filling\cite{Regnault09,Zozulya09,Lauchli10,Thomale10-1,Sterdyniak11-1,Thomale11,Chandran11,Sterdyniak11-2,Qi12},
quantum Hall bilayers at filling factor $\nu=1$\cite{Schliemann11},
the Kitaev model\cite{Yao10}, one dimensional quantum spin
systems\cite{Calabrese08,Fidkowski10,Thomale10-2,Pollmann10-1,Pollmann10-2,Turner11,Chiara12,Lepori13,Rao14,Lundgren15},
Hofstadter poblem\cite{Huang12, Schliemann13, Schliemann14},
interacting fermions on honeycomb lattice\cite{Assaad14} and bosonic
critical system in three dimensions\cite{Lemonik15}. As a result of
these studies, the ES depends on the chosen basis to partition the
many body Hilbert space. For systems with a bulk energy gap, the ES
obtained from some form of spatial cut contains information
reflecting the actual excitation spectrum of the systems at
issue.\cite{Thomale10-1, Lauchli10, Qi12,Lundgren13}. A large amount
of focus of the investigation of ES has been with a partition in
various systems like two-leg ladders and various bilayer systems,
where the edge comprises the entire remaining subsystem.
\cite{Poilblanc10,Cirac11,Peschel11,Lauchli11,Tanaka12,Lundgren12,Lundgren13,Chen13,Moudgalya15,Santos15}.
In this systems, broadly speaking, in the strong interlayer coupling
limit, the entanglement Hamiltonian is proportional to the
subsystems Hamiltonian. However, this is not the case in general and
their relation, depends on the couplings between subsystems.
In this paper we consider a model of non-interacting free fermions on a
bilayer honeycomb lattice where the presence of interlayer skew
hoppings breaks down the mentioned proportionality. Utilizing single particle
correlations of one layer, we obtain the entanglement Hamiltonian of
the system and present a relation between hopping parameters in
which, an exact correspondence is established between the ES of the
bilayer and energy spectrum of the monolayer (MES). We show that
this correspondence breaks down in the presence of trigonal warping
in the band structure of the bilayer. However, similar to the band
structure, the ES is symmetric with respect to the zero entanglement
as well as the trigonal warping appears on the ES. In the presence
of on-site energy difference, in spite of the symmetric energy
spectrum of bilayer the entanglement spectrum is completely
asymmetric.

In the second part of this paper we study the entanglement spectrum of bilayer honeycomb lattices in the presence of a uniform perpendicular magnetic field. We
demonstrate that in the absence of trigonal wrapping in energy spectrum, the ES similar to MES has a self-similar fractal structure, dubbed {\it Hofstadter
butterfly}. We also show that, the presence of an on-site energy
difference results in a transition from the so-called
Hofstadter butterfly to a tree-like picture.
\begin{figure*}
\centering
\includegraphics[width=65mm]{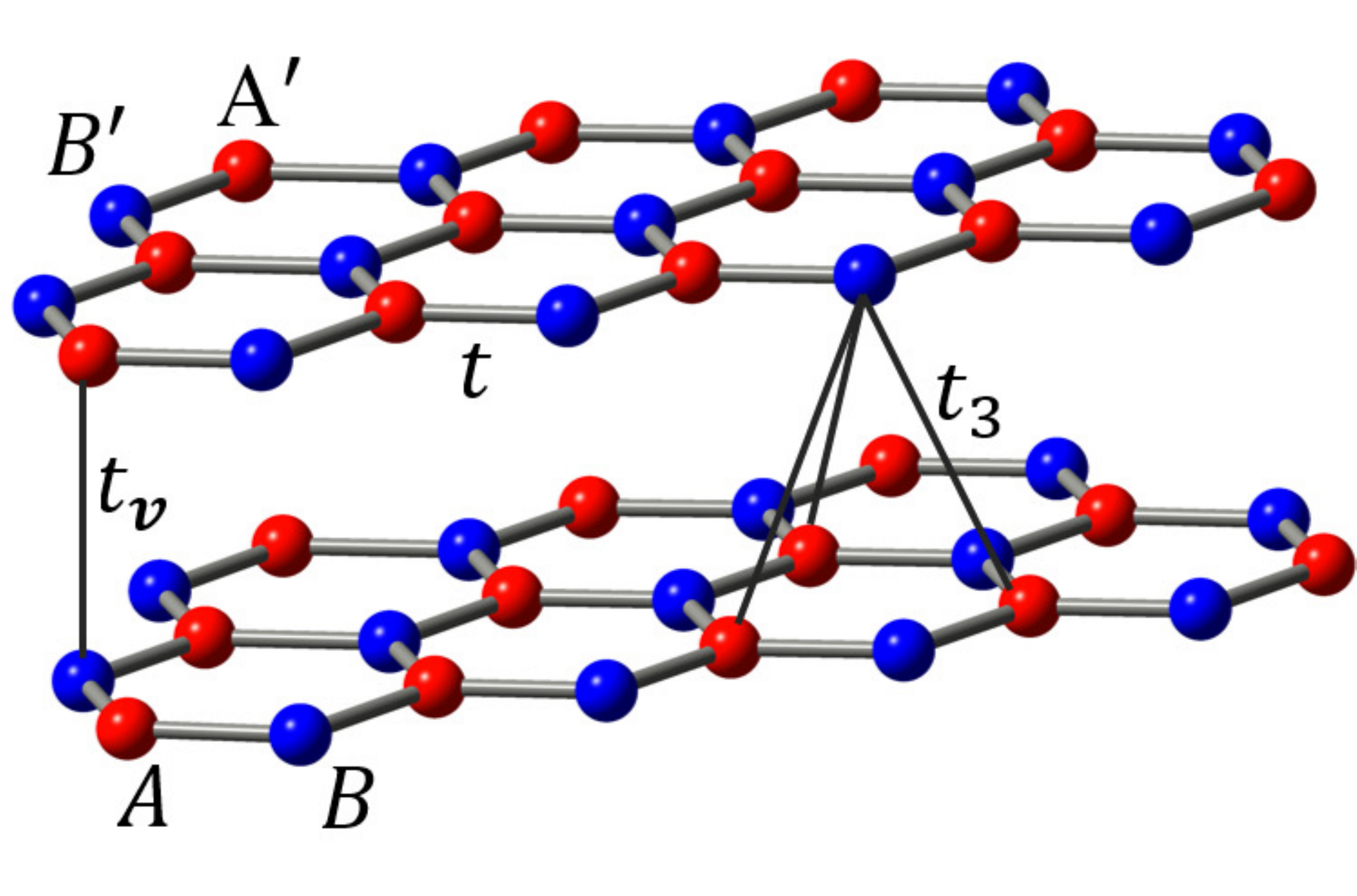}
\includegraphics[width=180mm]{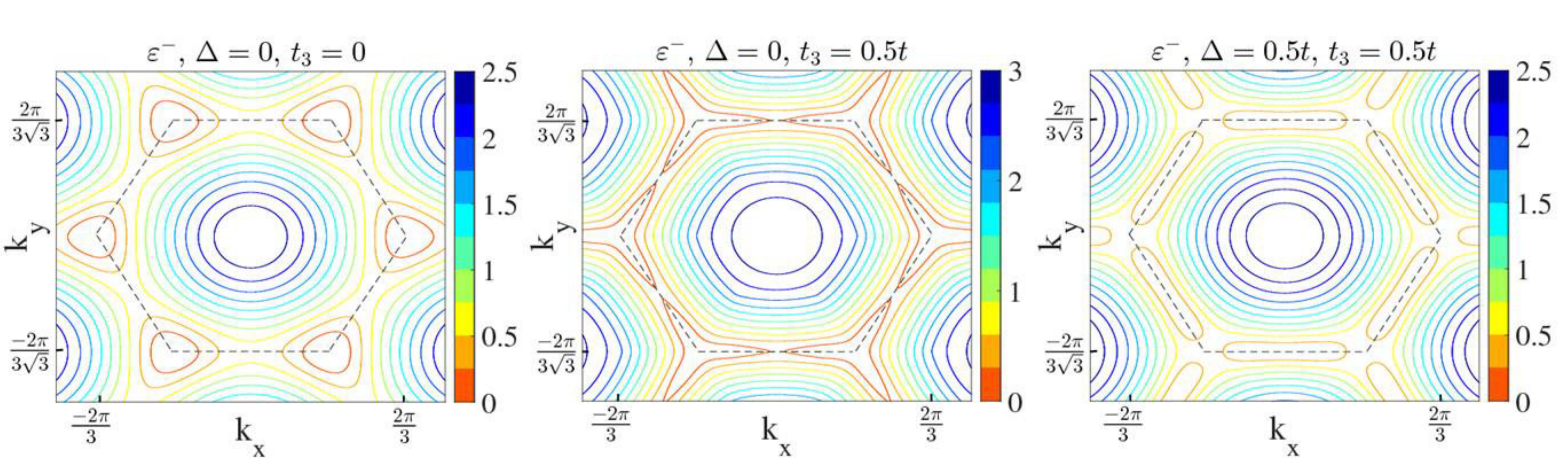}
\includegraphics[width=180mm]{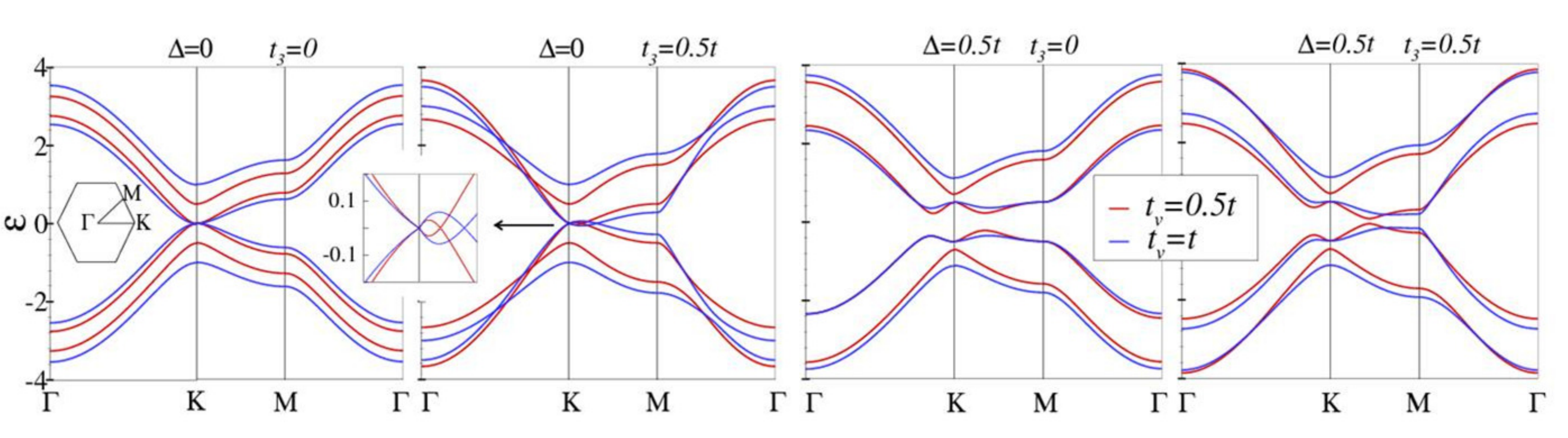}
\caption{(Color online) Top: Illustration of Bernal-stacked
fermionic bilayer honeycomb lattice with the indication of the
within layer and interlayer hopping parameters. Middle: contour plots of energy band $\e^-$ for
$t_v=t$. The dashed lines indicate the edge of the Brillouin zone. Colors show the amount of entanglement. In the
presence of interlayer skew hopping three additional Dirac points appear near a
given K point.  Bottom: the band structure for vertical hopping
parameters $t_v=0.5 t$ (red lines) and $t_v=t$ (blue lines). Two
left panels correspond to the energy spectrum in the absence of on-site
energy difference for skew hoppings $0$ and $0.5 t$. The inset
hexagon is the first Brillouin zone with labels of high symmetry
points. The enlarged inset plot shows effects of trigonal warping on
the band structure in the vicinity of zero energy at the edge of
Brillouin zone. Two right panels correspond to energy spectrum in
the presence of on-site energy difference,
$2\Delta$.} \label{bi-energy-band}
\end{figure*}

We have organized the rest of this paper as follows. In Sec.
\ref{sec:model}, we introduce the non-interacting fermionic model on
a  bilayer honeycomb lattice and obtain the energy spectrum of the
model in the presence of interlayer skew hoppings and on-site energy
differences. In Sec. \ref{sec:ent} using nonzero single particle
correlation functions we build the layer-layer entanglement
Hamiltonian. By diagonalizing this  Hamiltonian we obtain the
entanglement spectrum of the bilayer. In this section we investigate
in details the effects of trigonal warping and on-site energy
difference on the behavior of single particle correlations and their
direct relations with the entanglement spectrum of the system. In
Sec. \ref{sec:HoF} we investigate the effects of an external
perpendicular magnetic field on the energy band structure and
entanglement spectrum of bilayer. Finally in Sec. \ref{sec:con} we
present our conclusions and outlook.
\section{Bilayer honeycomb lattice model}\label{sec:model}

Bilayer honeycomb lattice consists of two stacked hexagonal
monolayers which in turn are made of two sublattices. We denote them
as $A$ and $B$ for the lower layer and $A^\prime$ and $B^\prime$
for the upper layer. In the Bernal stacked bilayer honeycomb lattice (BLH)
the $A$ and $B^\prime$ sublattices are directly situated above each other as
illustrated on the top panel of Fig. \ref{bi-energy-band}, however the sites on other two sublattices, B and $A^\prime$, don't have a counterpart
on the other layer that is directly above or below
them.

The tight binding Hamiltonian of the model of noninteracting free
fermions on BLH latices is given by \cite{bilayergraphene}:
\begin{eqnarray}
\no H&=&-\sum_\kv[t S_\kv(a_\kv^\dg b_\kv+h.c.)-\Delta (a_\kv^\dg a_\kv+b_\kv^\dg b_\kv)]~~~\\
\no&-&\sum_\kv[t S_\kv(a_\kv^{\prime\dg} b_\kv^\prime+h.c.)+\Delta (a_\kv^{\prime\dg} a_\kv^\prime+b_\kv^{\prime\dg} b_\kv^\prime)]\\
&+& \sum_\kv [ t_ v (b_\kv^\dg a_\kv^\prime +a_\kv^{\prime\dg} b_\kv)
-t_3 (S_\kv b_\kv^{\prime\dg} a_\kv + S^*_\kv a_\kv^\dg
b_\kv^\prime)],
\label{HBL-kspace}
\end{eqnarray}
where the summations run over wave vectors in hexagonal Brillouin zone. $a_\kv^\dg (a_\kv^{\prime\dg})$, $b_\kv^\dg
(b_\kv^{\prime\dg})$  and $a_\kv (a_\kv^\prime)$, $ b_\kv
(b_\kv^\prime)$ are fermionic creation and annihilation operators with wave vector $\kv$ in
lower (upper) layer on sublattices $A (A^\prime)$ and $B
(B^\prime)$, respectively.  $S_\kv= \sum_{\l=1}^3 e^{i\kv\cdot\vec{\delta}_\l}$ where $\vec{\delta}_\l$ are the positions
of three nearest $B$ sites relative to a given $A$ site, which is written as
\begin{equation}
\vec{\delta}_{1,2}=\frac
a2(1,\pm\sqrt{3}),~~~\vec{\delta}_3=a(-1,0),
\end{equation}
with $a$ the distance between two neighboring sites. The on-site energy difference between two layers
is denoted by $2\Delta$ which can be controlled by doping or an external electric field.
$t$ denotes the hopping within each layer
between sublattices and $\tv$ is vertical hopping between two
sublattices in different layers, $B$ and $A^{\prime}$. The parameter
$t_3$ is skew hopping between $A$ and $B^{\prime}$ which leads to
trigonal warping in the energy spectrum. All these hoppings
are illustrated in Fig. \ref{bi-energy-band}.
\begin{figure*}[t]
\centering
\includegraphics[width=180mm]{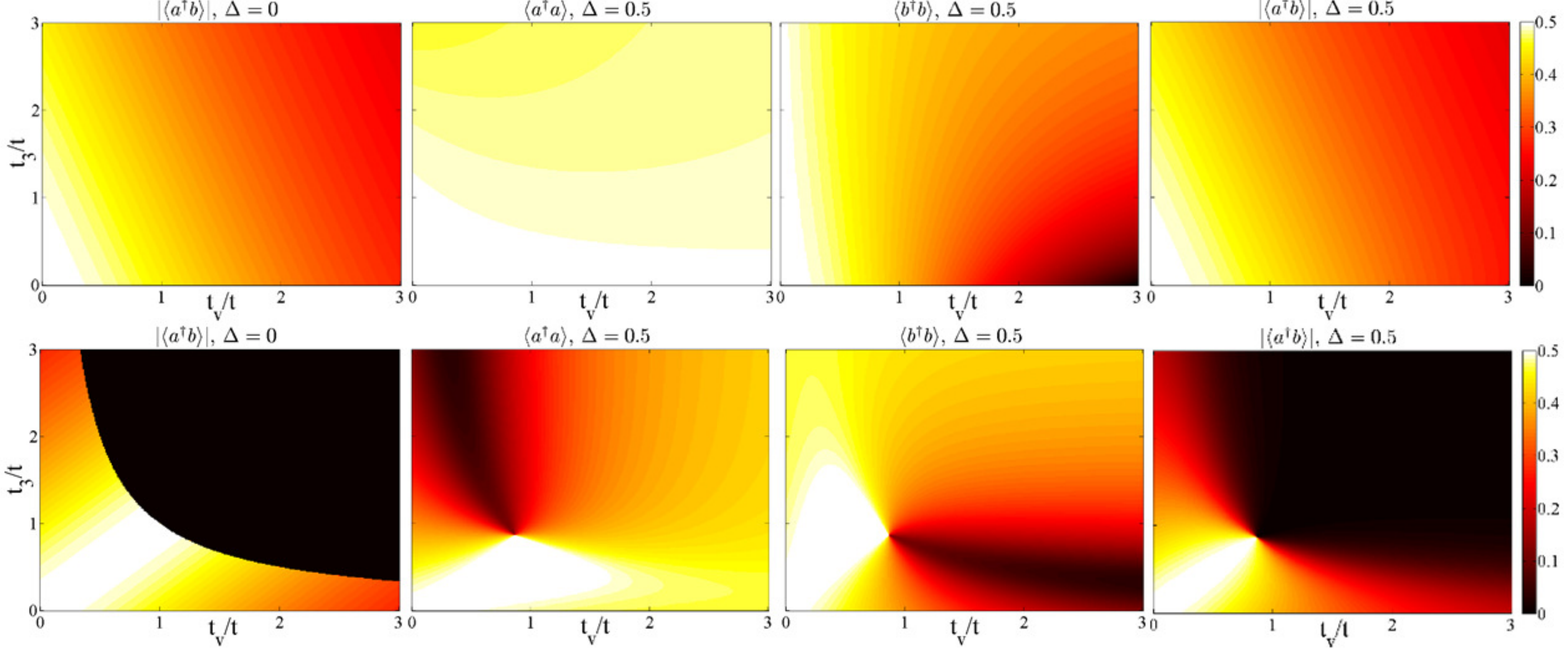}
\caption{(Color online) Single particle correlation functions versus
scaled hopping parameters $t_v/t$ and $t_3/t$ in the absence and
presence of on-site energy difference. Top: at $\Gamma$ point.
Bottom: at M point.} \label{correlation}
\end{figure*}

By diagonalizing the Hamiltonian (\ref{HBL-kspace}), the energy spectra of the bilayer are readily obtained as $\e_\kv=\pm\e^{\pm}$, with
\begin{widetext}
\begin{eqnarray}
\varepsilon^\pm=\sqrt{\Delta^2 + \frac 12\left(\tv^2 +(t_3^2 + 2t^2)|S_\kv|^2
\pm\sqrt{4t^2|S_\kv|^2(4\Delta^2 + \tv^2 + t_3^2|S_\kv|^2) +(\tv^2 - t_3^2|S_\kv|^2)^2+ 8t^2t_3\tv\Re(S_\kv^3)}\right)},
\label{energyspect}
\end{eqnarray}
\end{widetext}
where $\Re$ denotes real part. The band structure is symmetric with
respect to zero energy ($\e_{\rm F}=0$), and the separation of
branches depends on the hopping parameters and on-site energy
difference (see Fig. \ref{bi-energy-band}). In the absence of
on-site energy difference, at zero skew hopping the band structure
is rotationally symmetric for an area around a K point and shows a
split into four massive bands, with the two inner ones ($\pm\e^-$)
touching at zero energy and the two outer branches ($\pm\e^+$)
separating from the inner ones by gap $t_v$. In the presence of skew
hoppings, level crossings between $\e_+$ and $\e_-$ and between
$-\e_+$ and $-\e_-$ occur in $\Gamma$-K direction and therefore the
low energy physics of the system is captured by the two new inner
bands $\pm\e^+$. At the edge of entire Brillouin zone these two
branches touch at zero energy at an additional Dirac point where
they have linear dispersion (Fig. \ref{bi-energy-band}, the bottom
panels, inset plot). The appearance of three additional Dirac points
around a given K point due to the skew hoppings is known as trigonal
warping which is shown by the contour plots in Fig.
\ref{bi-energy-band}.

The level crossings in the band structure
are removed by on-site energy difference. The on-site energy
difference creates separation $2\Delta$ between inner branches
and opens a band gap at the edge of Brillouin zone.

\section{Entanglement Spectrum of Bilayer Honeycomb Lattice}\label{sec:ent}

Before we start the discussions on layer-layer ES of the BLH
lattice, it is worthwhile to review briefly some mathematical
details concerning the ES of fermionic hopping models on two-leg
ladders\cite{Peschel03-1,Peschel11} or bilayer square
lattices\cite{Schliemann13}. In these systems the Hamiltonian of
each layer is generally written in the following diagonal form
\begin{equation}
H_\l = \sum_\kv\e_{\l\kv} c_{\l\kv}^\dg c_{\l\kv},
\end{equation}
where $c_{\l\kv}^\dg$ and $c_{\l\kv}$ are fermionic creation and annihilation operators of single particle states with eigenvalues
$\e_{\l\kv}$. For systems with opposite dispersion in both
subsystems {\it i.e.}, $\e_{1\kv}=-\e_{2\kv}$, the entanglement
Hamiltonian is readily obtained by tracing out subsystem $2$ as:
\begin{equation}
{\cal H}_{ent}= \sum_\kv \e_\kv^e c_{1\kv}^\dg c_{1\kv},
\end{equation}
where $\e_\kv^e$ denotes the entanglement spectrum which is given by
$\e_\kv^e=2{\rm arctanh}(2\braket{c_{1\kv}^\dg c_{1\kv}})$, with the
single particle correlation $\braket{c_{1\kv}^\dg c_{1\kv}}$ on the
ground state of the composite
system\cite{Peschel03-1,Peschel09,Vidal03,Latorre09}.

The non-interacting free fermions model on BLH lattice
(\ref{HBL-kspace}) is somewhat different from the models considered
so far, in that the Hamiltonian of layers does not have a diagonal
form. We propose the following way for determining the entanglement
Hamiltonian of this system. In the first step, we obtain non-zero
single particle correlations of lower layer's fermionic operators
($a$ and $b$) on the ground state of the composite system. In the
second step, we build a fictitious entanglement Hamiltonian in terms
of the single particle operators with non-zero correlations.

By diagonalizing the Hamiltonian (\ref{HBL-kspace}) we find out that the single particle correlations:
\begin{equation}
\label{non-zero-corr}
\bra{\psi}a_\kv^\dg a_\kv\ket{\psi},
\bra{\psi}b_\kv^\dg b_\kv\ket{\psi},
\bra{\psi}a_\kv^\dg b_\kv\ket{\psi},
\bra{\psi}b_\kv^\dg a_\kv\ket{\psi},
\end{equation}
are non-zero where $\ket{\psi}$ denotes ground state of composite system.
Employing these non-zero correlations we build
the entanglement Hamiltonian as:
\begin{equation}
{\cal H}_{ent}=\sum\limits_\kv u_\kv^a a_\kv^\dg
a_\kv+u_\kv^bb_\kv^\dg b_\kv+(v_\kv a_\kv^\dg b_\kv+h.c.),
\label{NDEntH}
\end{equation}
where the unknown $\kv$-dependent coefficients $u_\kv^a$, $u_\kv^b$, and
$v_\kv$ could be obtained in terms of the nonzero single particle correlation functions (\ref{non-zero-corr}),
however we will see that their explicit forms are not necessary to achieve the ES.
By defining
\begin{equation}\label{ent-spect}
\xi_\kv^\pm=\frac 12\left(u_\kv^a-u_\kv^b\pm\sqrt{(u_\kv^a-u_\kv^b)^2+4|v_\kv|^2}\right),
\end{equation}
and employing the following unitary transformation:
\begin{eqnarray}
&&\no
a_\kv=\frac{\xi_\kv^-/v_\kv^*}{\sqrt{1+(\xi_\kv^-/|v_\kv|)^2}}\a+
\frac{\xi_\kv^+/v_\kv^*}{\sqrt{1+(\xi_\kv^+/|v_\kv|)^2}}\b,\\
&&b_\kv=\frac{1}{\sqrt{1+(\xi_\kv^-/|v_\kv|)^2}}\a+
\frac{1}{\sqrt{1+(\xi_\kv^+/|v_\kv|)^2}}\b,\label{unitary-trans}
\end{eqnarray}
where, $\a$ and $\b$ are two new fermionic operators, the entanglement
Hamiltonian (\ref{NDEntH}) is readily diagonalized as
\begin{eqnarray}
{\cal H}_{ent}^{d}=\sum_\kv(\xi_\kv^+\alpha_\kv^\dg\a+\xi_\kv^-\beta_\kv^\dg\b).
\end{eqnarray}
Here, $\xi_\kv^\pm$ are the entanglement spectra which could be written in terms of $n_\kv^+=\braket{\alpha_\kv^\dg\a}$ and $n_\kv^-=\braket{\beta_\kv^\dg\b}$ as
\begin{equation}
\xi_\kv^\pm=2{\rm arctanh}(2n_\kv^\pm).
\label{ent-spect-gen1}
\end{equation}
It is now needed to find $n_\kv^\pm$ on the ground state of the
composite system. By substituting $\xi^\pm$ from Eq. (\ref{ent-spect}) into the unitary
transformations (\ref{unitary-trans}) we readily obtain them after simple manipulations in terms of the non zero correlations (\ref{non-zero-corr})
as follows:
\begin{eqnarray}
\no n_\kv^\pm&=&\frac 12\bigg(\braket{a_\kv^\dg a_\kv}+\braket{b_\kv^\dg b_\kv}\\
&\pm&\sqrt{(\braket{a_\kv^\dg a_\kv}-\braket{b_\kv^\dg
b_\kv})^2+4|\braket{a_\kv^\dg b_\kv}|^2}\bigg). \label{ent-spect-gen2}
\end{eqnarray}
In the absence of on-site energy differences, $\braket{a_\kv^\dg
a_\kv}$ and $\braket{b_\kv^\dg b_\kv}$ are identically equal to
$1/2$ which implies the existence of the symmetry of single particle
distributions on sublattices $A$ and $B$ of lower layer in the
ground state of composite system. The single particle correlations
$|\braket{a_\kv^\dg b_\kv}|$, however, depend on $\kv$ vector as
well as the hopping parameters. At a given K point they are zero,
whereas their behavior at other high symmetry points strongly
depends upon the hopping parameters. At $\Gamma$ point, for small values
of $t_v/t$ and $t_3/t$ they are almost $1/2$ and increasing the
interlayer couplings begin to progressively destroy them and for
large enough values of interlayer hoppings they will be destroyed
(see upper left panel of Fig. \ref{correlation}). At a given M point
they fall sharply down to zero by increasing the interlayer
couplings (see lower left panel of Fig. \ref{correlation}).
\begin{figure}
\centering
\includegraphics[width=89mm]{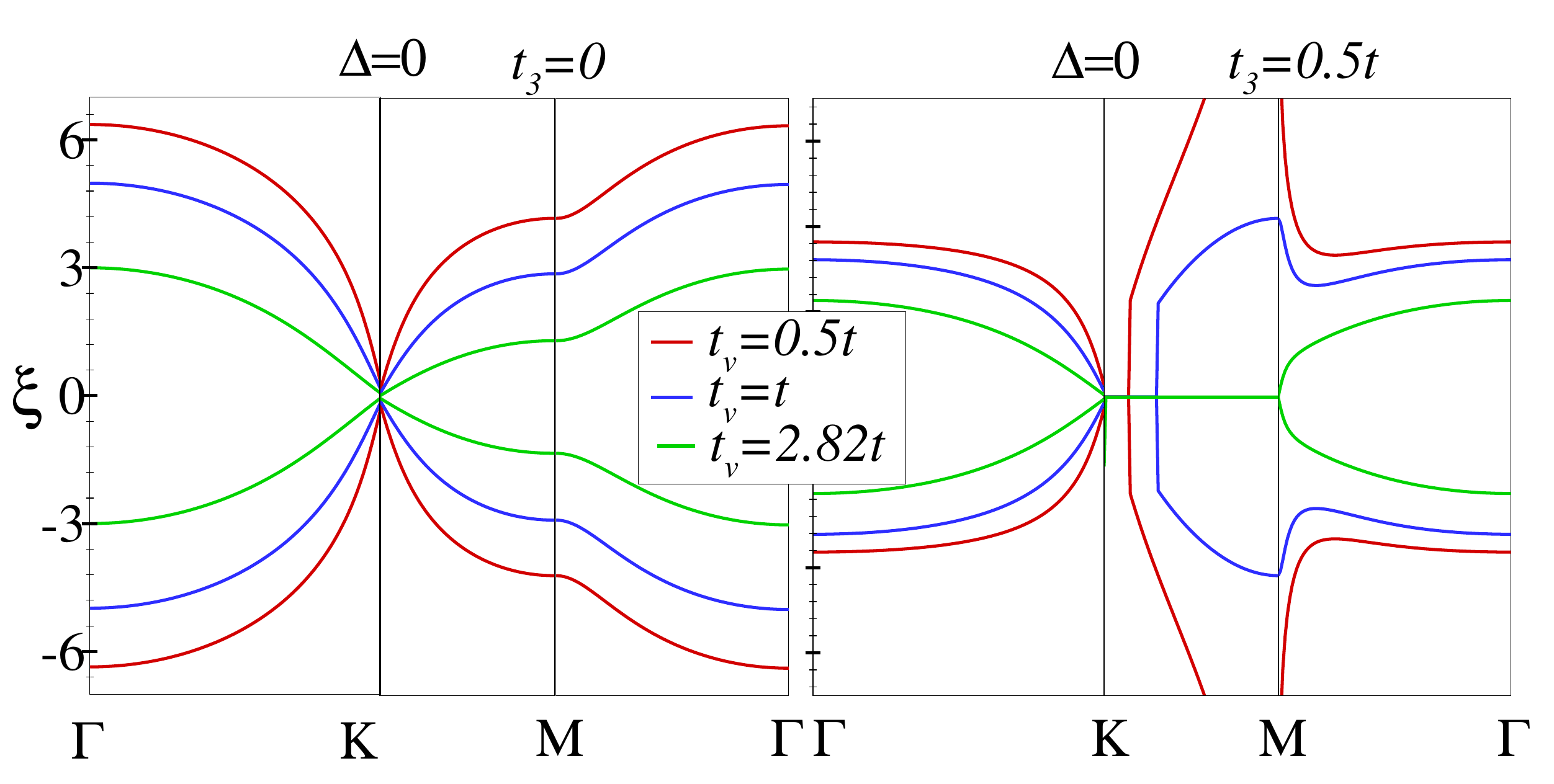}
\includegraphics[width=60mm]{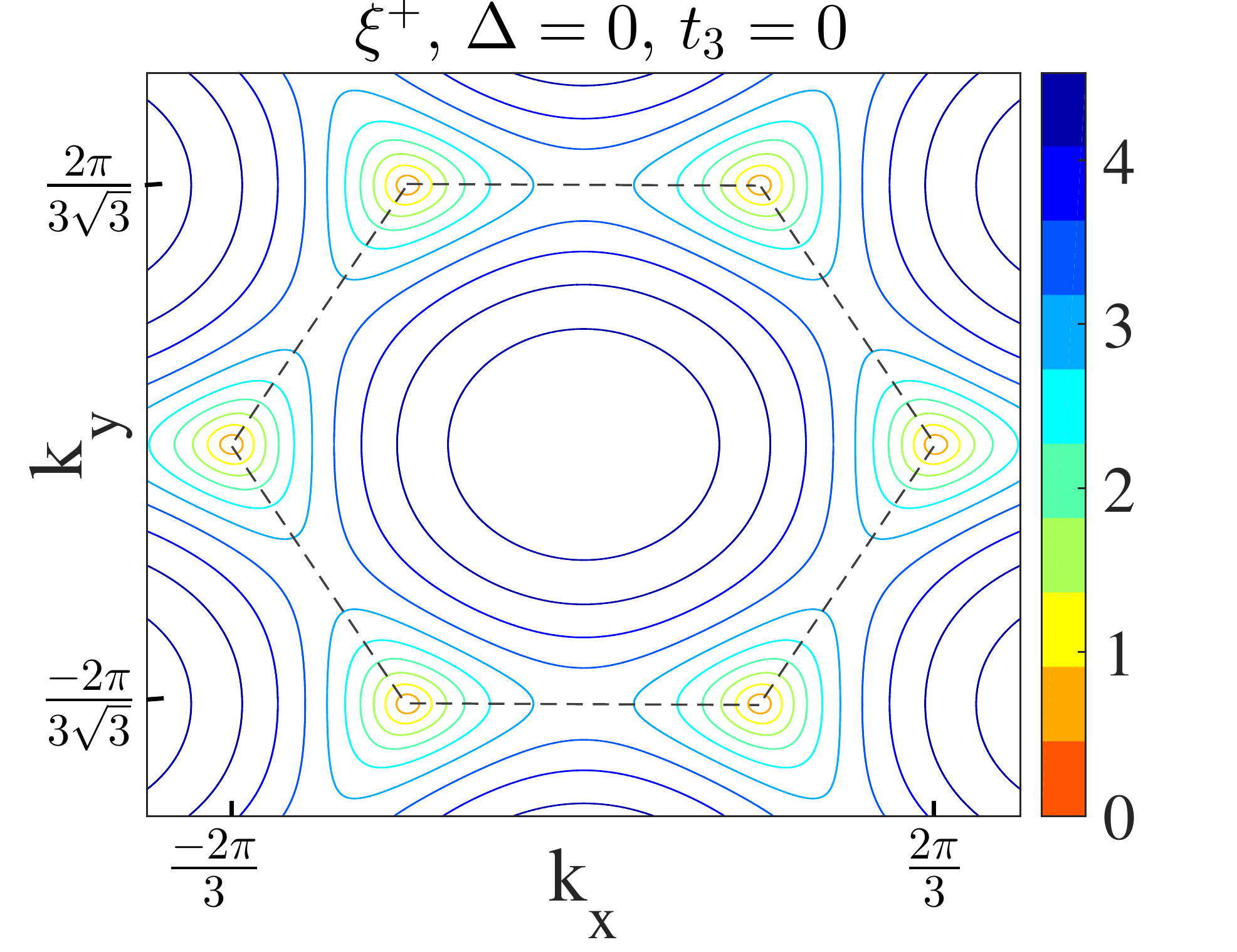}
\includegraphics[width=60mm]{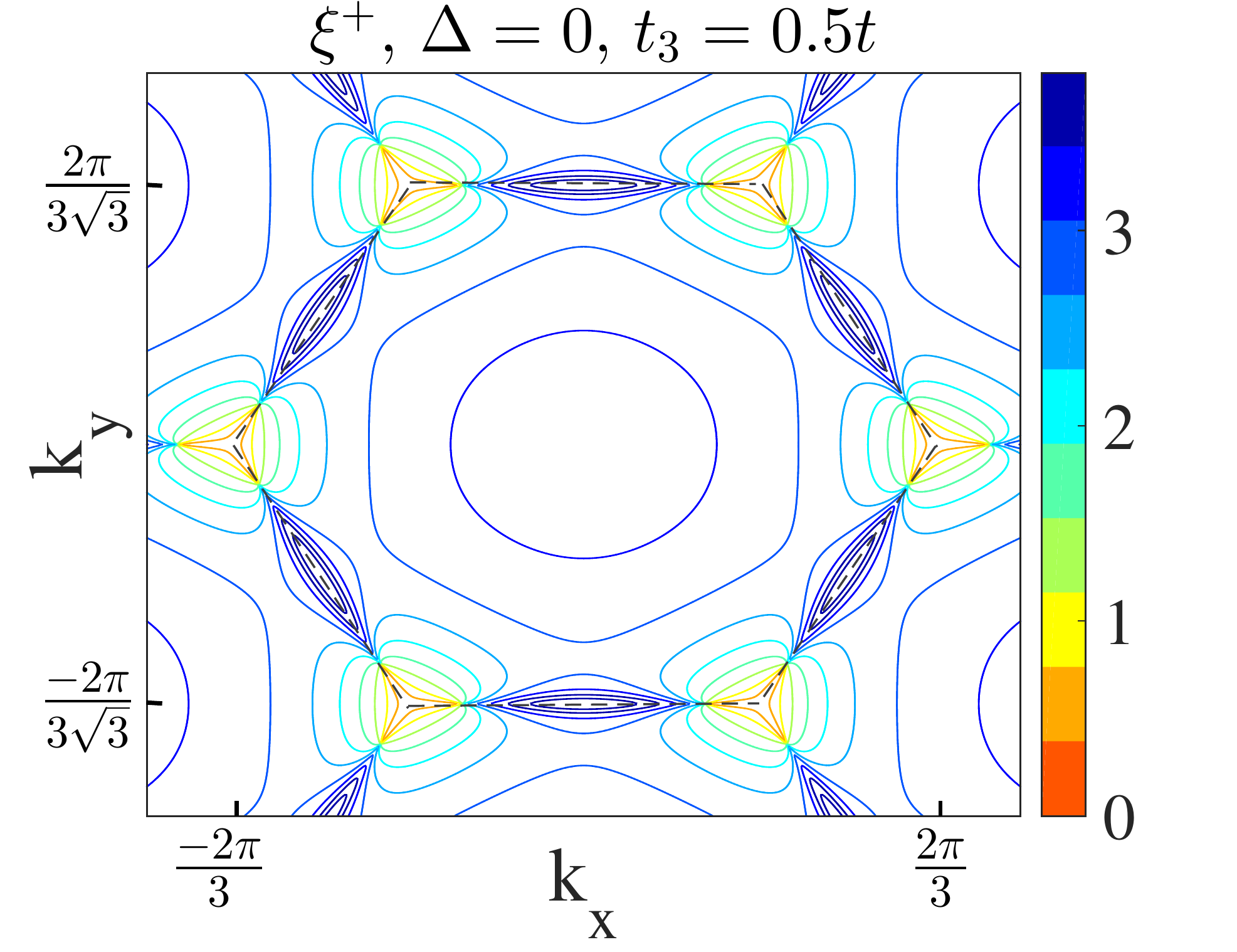}
\caption{(Color online) Entanglement spectrum of Bernal-stacked
fermionic bilayer honeycomb lattice in the absence of on-site energy
difference. Top: entanglement spectra versus $\kv$ vector for
different values of vertical hopping, in the absence (left panel)
and presence (right panel) of interlayer skew hoppings. The green
line is both the energy spectrum of a monolayer honeycomb lattice
and the ES of bilayer at $t_v\cong 2.82 t$. Bottom: contour plots of
$\xi_\kv^+$. The dashed lines indicate the edge of the Brillouin
zone. Similar to the energy spectrum of BLH lattice, trigonal
warping also appear near K point in the ES.} \label{entspect-d0}
\end{figure}

In the absence of on-site energy difference ($\Delta=0$), the entanglement spectra (\ref{ent-spect-gen1}) are simplified to:
\begin{equation}
\xi_\kv^+=-\xi_\kv^-=-2{\rm arctanh}(2|\braket{a_\kv^\dg b_\kv}|),\label{ent-spect-d0}
\end{equation}
where, similar to the band structure, they are symmetric with
respect to the level with zero entanglement ($\xi=0$) and their
separation depends on the interlayer couplings $t_v/t$ and $t_3/t$ (see
Fig. \ref{entspect-d0}). In the absence of trigonal warping on the energy spectrum of the bilayer,
the entanglement spectra are rotationally symmetric for an area around a K point and shows a split into two massless bands, touching at
$\xi=0$. At a particular relevance between interlayer and within
layer hoppings, {\it i.e.} at $t_3=0$ and $t_v\cong 2.82 t$, an
exact correspondence is established between the entanglement spectra
of BLH lattice and monolayer energy spectra which means that the
layer-layer ES of BLH lattice reflects the edge state properties of
the system, perfectly.

Interlayer skew hoppings considerably affect the ES of the system.
As a consequence of trigonal warping some degeneracy and one
discontinuity are shown to take place at the edge of entire
Brillouin zone, {\it i.e.} in K-M direction, due to the level
touching at zero energy on band structure of BLH lattice (see Fig.
\ref{entspect-d0}, plot $\Delta=0$ and $t_3=0.5 t$). In the presence
of trigonal warping, in general, there is no exact correspondence
between entanglement spectra of BLH lattice and the energy spectrum
of a monolayer honeycomb lattice , however, in $\Gamma$-K direction
for particular relevances between interlayer and within layer
hoppings, {\it e.g.} for $t_3\sim 0.1 t$ and $t_v\sim 2.51 t$, the
behavior of the ES versus $\kv$ are remarkably similar to the MES.
\begin{figure}[h]
\centering
\includegraphics[width=87mm]{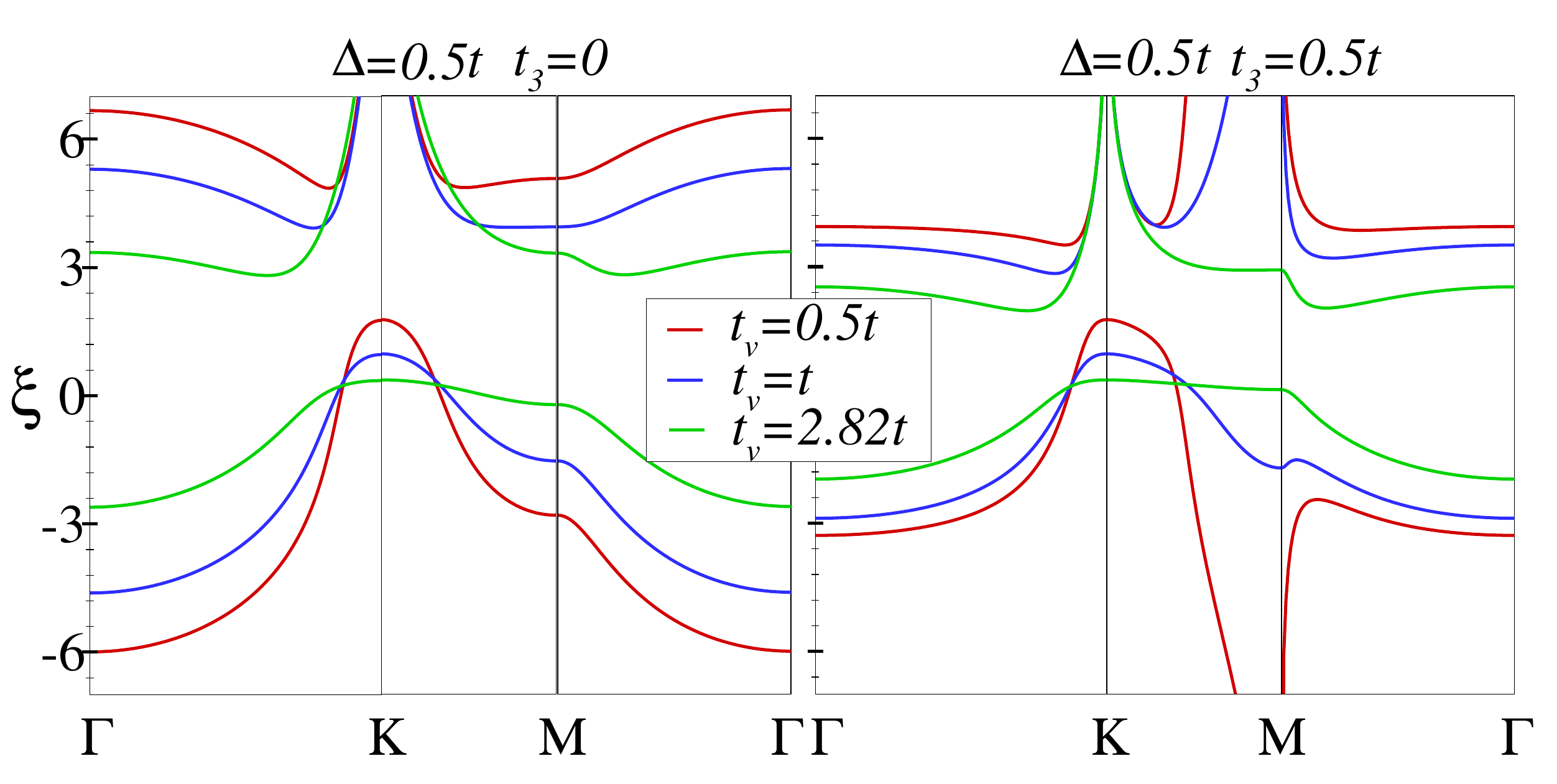}
\includegraphics[width=60mm]{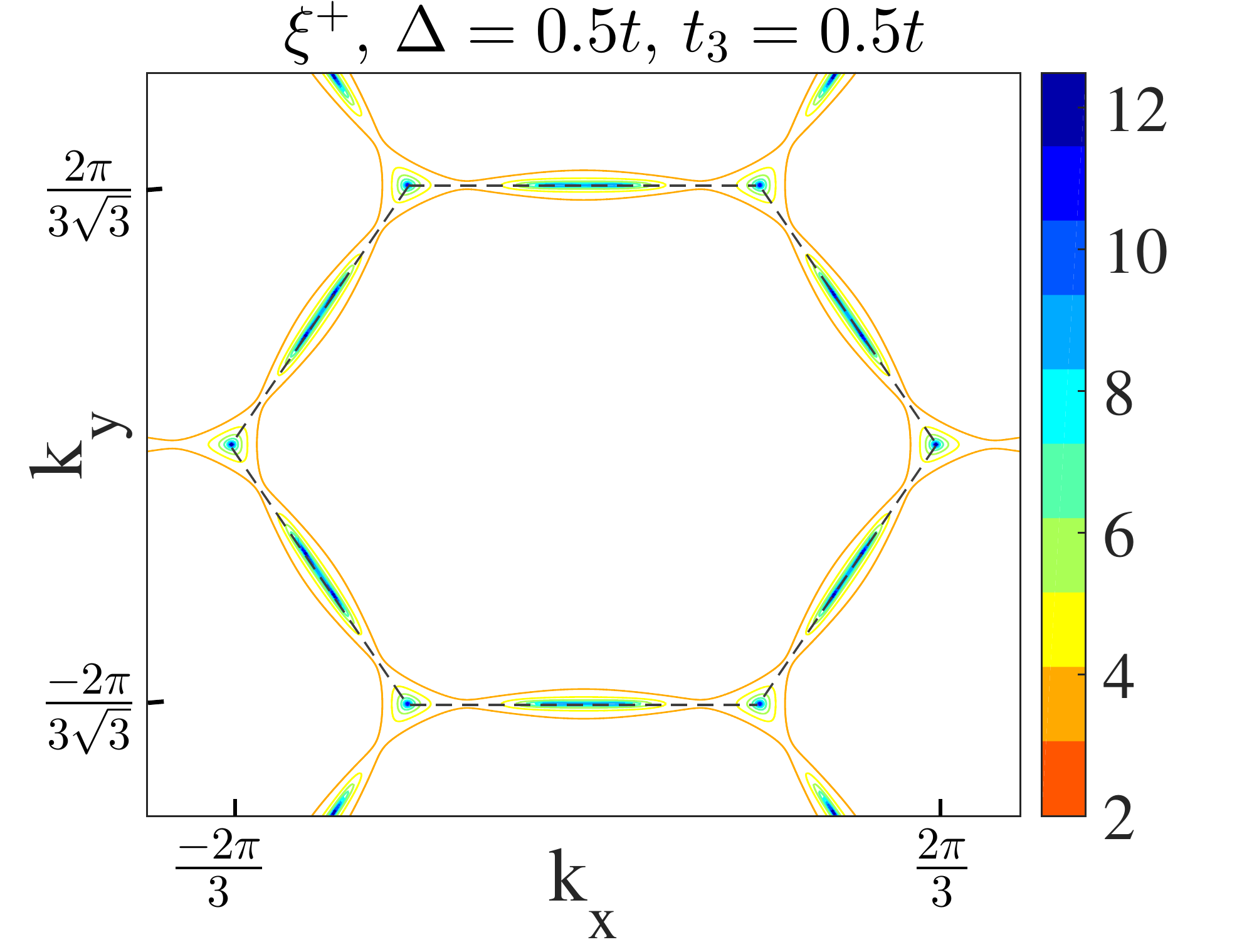}
\includegraphics[width=60mm]{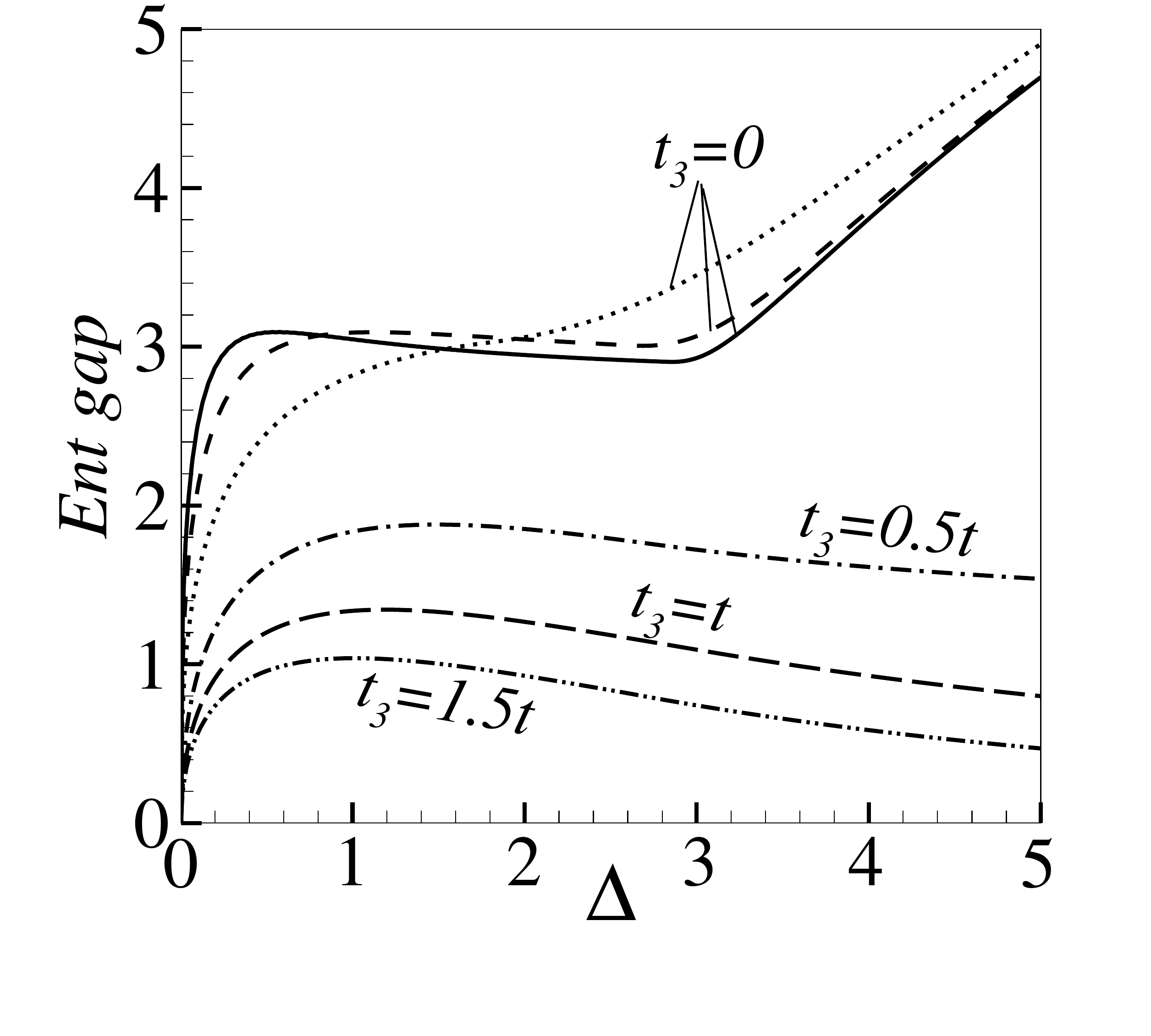}
\caption{(Color online) Entanglement spectrum and entanglement gap of Bernal-stacked fermionic bilayer honeycomb lattice in the presence of on-site energy difference.
Top: entanglement spectra versus $\kv$ vector for different values of vertical hopping, in the absence (left panel) and presence (right panel) of trigonal warping.
Middle: contour plot at the presence of on-site energy difference.
Bottom: the indirect entanglement gap versus on-site energy difference.
The solid line is for $t_v=0.5 t$, the dashed line is for $t_v=t$ and the others are for $t_v=2.82 t$.} \label{entspect-d05}
\end{figure}

The on-site energy difference ($\Delta\neq 0$) causes the
correlations $\braket{a_\kv^\dg a_\kv}$ and $\braket{b_\kv^\dg
b_\kv}$ behave differently with varying interlayer couplings (see
Fig. \ref {correlation}) which implies that the on-site energy
difference breaks the symmetry of single particle distributions on
sublattice $A$ and $B$ of the lower layer.  The different behaviors
of the single particle correlations, accompany an asymmetry in
entanglement spectra with respect to $\xi=0$ (see the upper panel of Fig. \ref{entspect-d05}). As a result of this
symmetry breaking, the degeneracy of the entanglement spectra at
$\xi=0$ is shown to be removed and consequently an {\it indirect
entanglement gap} (IEG) is opened. Variations of this IEG with increasing
the on-site energy difference is shown in lower panel of Fig.
\ref{entspect-d05}. In the absence of the trigonal warping it is
obviously seen that the IEG increases by $\Delta$, except for weak
interlayer vertical hoppings, where the IEG have very small
decrease. In the presence of trigonal warping, IEG is found to be
increased with a small value of $\Delta$, however more increasing of
the on-site energy decreases the IEG progressively and for
particular value of interlayer coupling the gap is exactly zero
where the two branches overlap indirectly.

\begin{figure}[t]
\centering
\includegraphics[width=70mm]{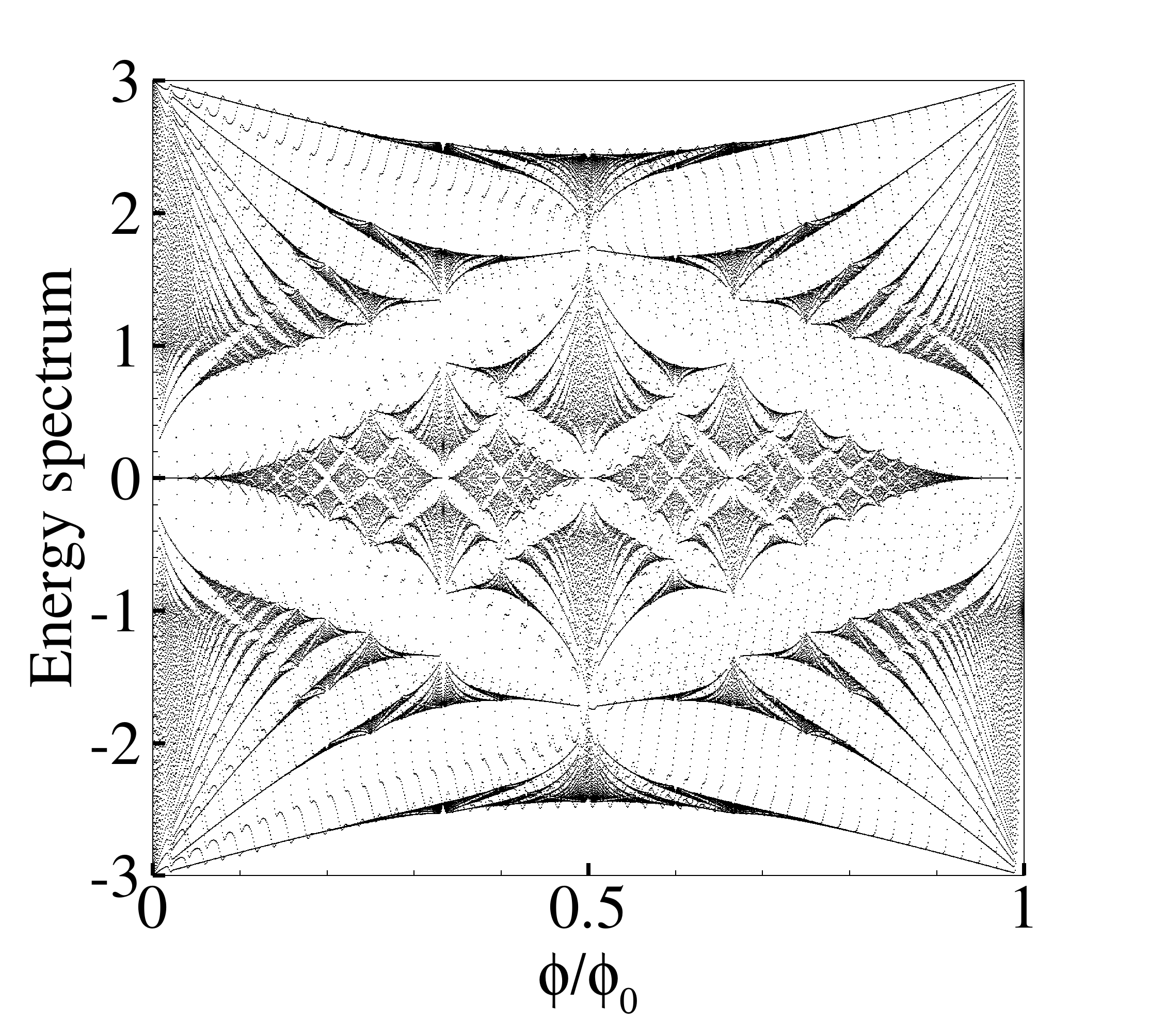}
\caption{Energy spectrum of a monolayer honeycomb lattice in the presence of a uniform magnetic field versus reduced magnetic flux $\phi/\phi_0=p/q$ where $0\leq p\leq q=51$.}
\label{HoFmono}
\end{figure}
\begin{figure*}
\centering
\includegraphics[scale=0.55]{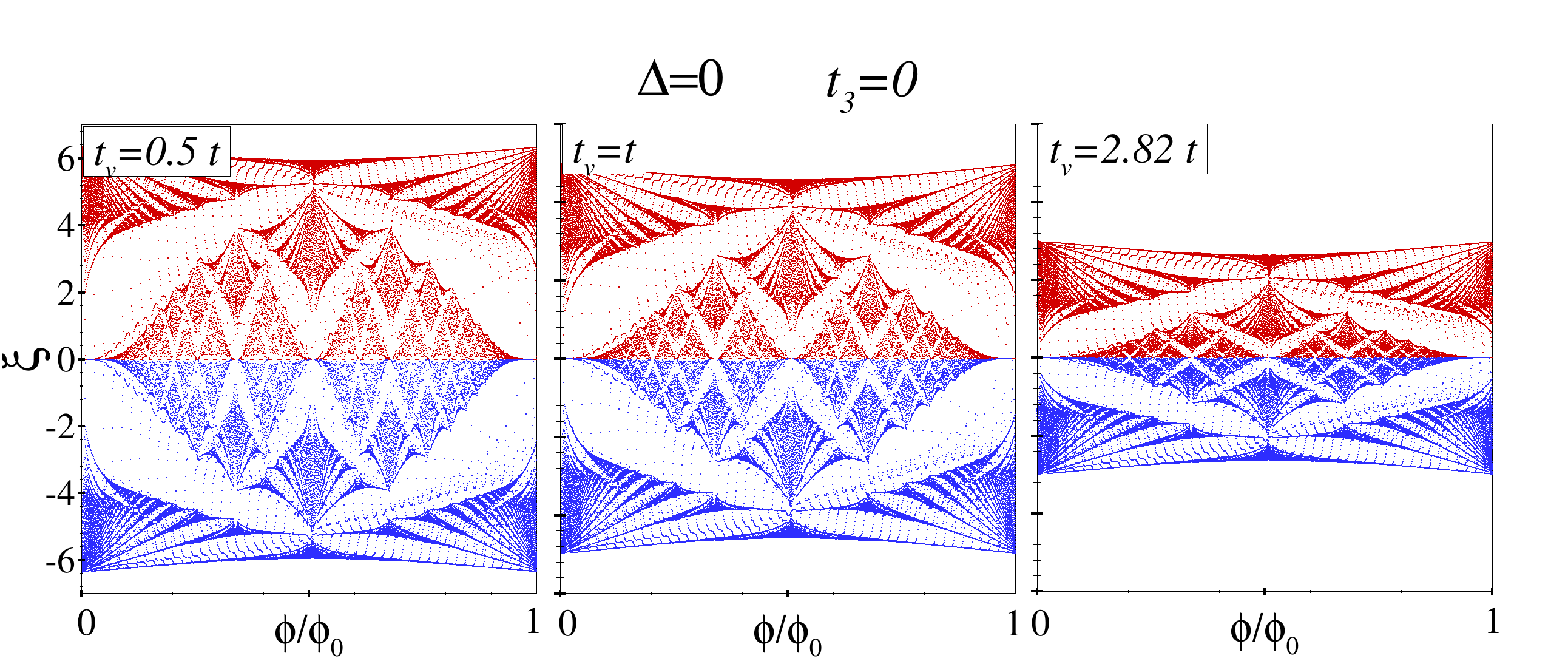}
\includegraphics[scale=0.6]{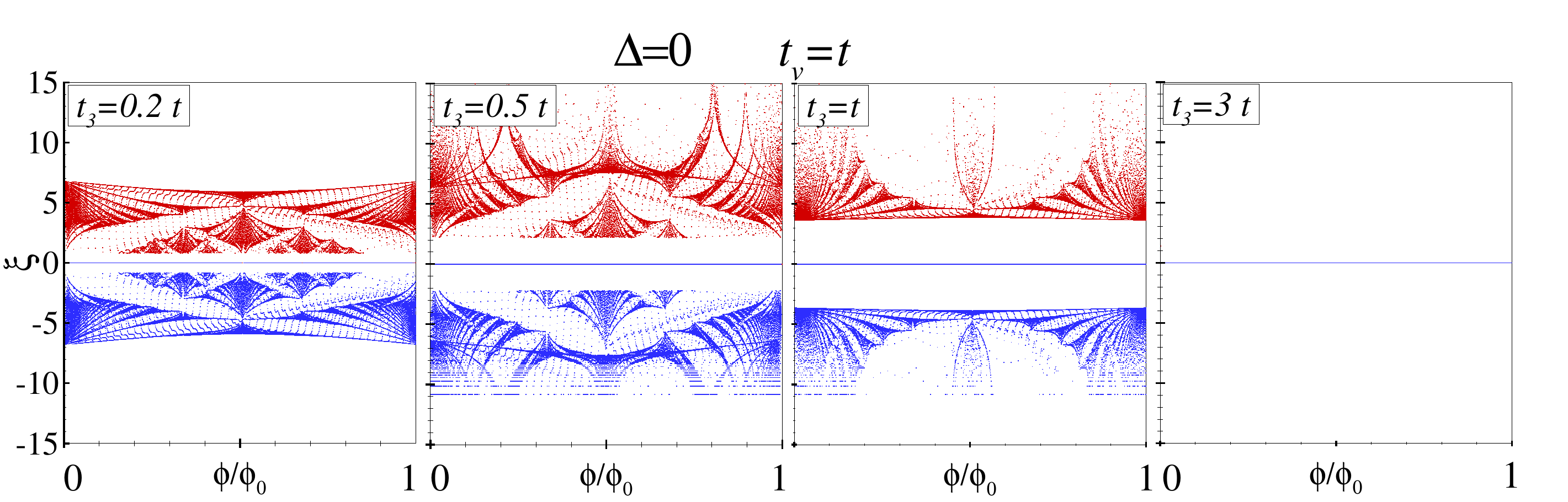}
\caption{Entanglement spectrum of BLH lattice versus reduced magnetic flux, in the absence of on-site energy difference, for $0\leq p\leq q=51$.
Top: symmetric self similar structure of entanglement spectrum in the absence of trigonal warping.
Bottom: asymmetric spectrum due to the effects of trigonal warping at $t_v=t$. For strong interlayer skew hoppings full collapse to zero entanglement is happened.}
\label{hofd0}
\end{figure*}

\section{Hofstadter bilayer honeycomb lattice model}\label{sec:HoF}

In this section we investigate the effects of an external uniform
magnetic field $\vec{B}=(0,0,B)$ on the entanglement spectrum of the
non-interacting free fermions model on the bilayer honeycomb
lattice. The effects of a uniform magnetic field, perpendicular to
the two dimensional fermionic lattices, could be included by
modifying the hopping parameters via the Peierls
substitution\cite{Peierls} as $t_{ij} =t e^{iA_{ij}}$, such that the
magnetic flux $\phi$ through each unit cell is a rational multiple
$p/q$ of the magnetic flux quantum, $\phi_0$. The gauge field
$A_{ij}$ can be made periodic by constructing a magnetic super cell
of $q$ structural unit cells of the lattice. For a lattice with an
$r$ element basis, this results in $qr$ energy sub-bands which in
general do not cross. Plotting these energies as a function of
$\phi/\phi_0=p/q$ yields to a self similar fractal structure, known
as Hofstadter butterfly\cite{Azbel, Hofstadter}.

The actual equations, that need to be solved to obtain the energy
spectrum, are known as Harper's equation which is the energy
eigenvalue equation for the Hamiltonian matrix\cite{Rammal}. For
non-interacting free fermions model on monolayer honeycomb lattice
the Harper's equation is equivalent to a $q\times q$ eigenvalue
problem. Solving this equation, the energy spectrum versus reduced
magnetic flux $\Phi=\phi/\phi_0$ has a self similar butterfly
structure which is shown in Fig. \ref{HoFmono}.

For fermionic model on the BLH lattice, Harper's equation is
equivalent to the following $4q\times 4q$ eigenvalue problem
\begin{equation}
H_\kv\Psi_\kv=\e_\kv\Psi_\kv,
\end{equation}
where $\Psi_\kv$ is the $4q$-component spinor
\begin{eqnarray}
\no\Psi_\kv&=&\bigg(\eta_{0\kv},\dots,\eta_{q-1\kv},\zeta_{0\kv},\dots,\zeta_{q-1\kv},\\
\no&&\gamma_{0\kv},\dots,\gamma_{q-1\kv},\delta_{0\kv},\dots,\delta_{q-1\kv}\bigg)^T,
\end{eqnarray}
and
\begin{equation}
H(\kv)=\left(
\begin{matrix}
\Delta\Iv_{q \times q} & th_\kv & 0 & t_3h^*_\kv\\
th^*_\kv & \Delta\Iv_{q \times q} & t_v\Iv_{q \times q} & 0 \\
0 & t_v\Iv_{q \times q} & -\Delta\Iv_{q \times q} & th_\kv \\
t_3h_\kv & 0 & th^*_\kv & -\Delta\Iv_{q \times q}
\end{matrix}
\right).
\label{BiHof1}
\end{equation}
Here, $\Iv_{q\times q}$ is the $q\times q$ unit matrix and $h_\kv$
is the following $q\times q$ matrix
\begin{equation}
h_\kv=\left(
\begin{matrix}
        x_{0\kv} & y_{0\kv} & 0 & \dots & z_\kv \\
        y_{0\kv}^* & x_{1\kv} & y_{1\kv} & \dots & 0  \\
         \vdots &  \ddots & \ddots & \ddots   & \vdots \\
         0  &  & \ddots  & \ddots  &  y_{q-1\kv} \\
        z_\kv^* & 0  & \ldots & y_{q-1\kv}^* & x_{q-1\kv}
\end{matrix}
\right),
\label{Hof}
\end{equation}
where,
\begin{eqnarray}
\no x_{m\kv}&=&2\cos(2\pi m\Phi+\frac{\sqrt 3}{2} k_y),\\
\no y_{m\kv}&=&1+e^{i(2\pi m\Phi+ \frac{\sqrt 3}{2} k_y)},\\
\no z_\kv&=&e^{-iqk_x} +e^{i(2\pi \Phi - \frac{\sqrt 3}{2} k_y - \frac{1}{2} qk_x)},
\end{eqnarray}
and $m=0,1,\dots,q-1$. The diagonalization of the Hamiltonian
(\ref{BiHof1}), in general, requires numerics, however explicit
analytical results are possible for the special values of the
magnetic flux $\phi$. The diagonalization of the Hamiltonian
(\ref{BiHof1}) is indeed reduced to the diagonalization of the
following $4\times 4$ Hamiltonian:
\begin{equation}
H_{m\kv}=\left(
\begin{matrix}
\Delta & t\tilde{\e}_{m\kv} & 0  & t_3\tilde{\e}_{m\kv} \\
t\tilde{\e}_{m\kv} & \Delta & t_v  & 0 \\
0 & t_v & -\Delta & t\tilde{\e}_{m\kv} \\
t_3\tilde{\e}_{m\kv} & 0 & t\tilde{\e}_{m\kv} & -\Delta
\end{matrix}
\right),
\label{BiHof2}
\end{equation}
where $\tilde{\e}_{m\kv}$ is $m$-th eigenvalue of $h_\kv$.
\begin{figure*}[t]
\centering\noindent
\includegraphics[scale=0.6]{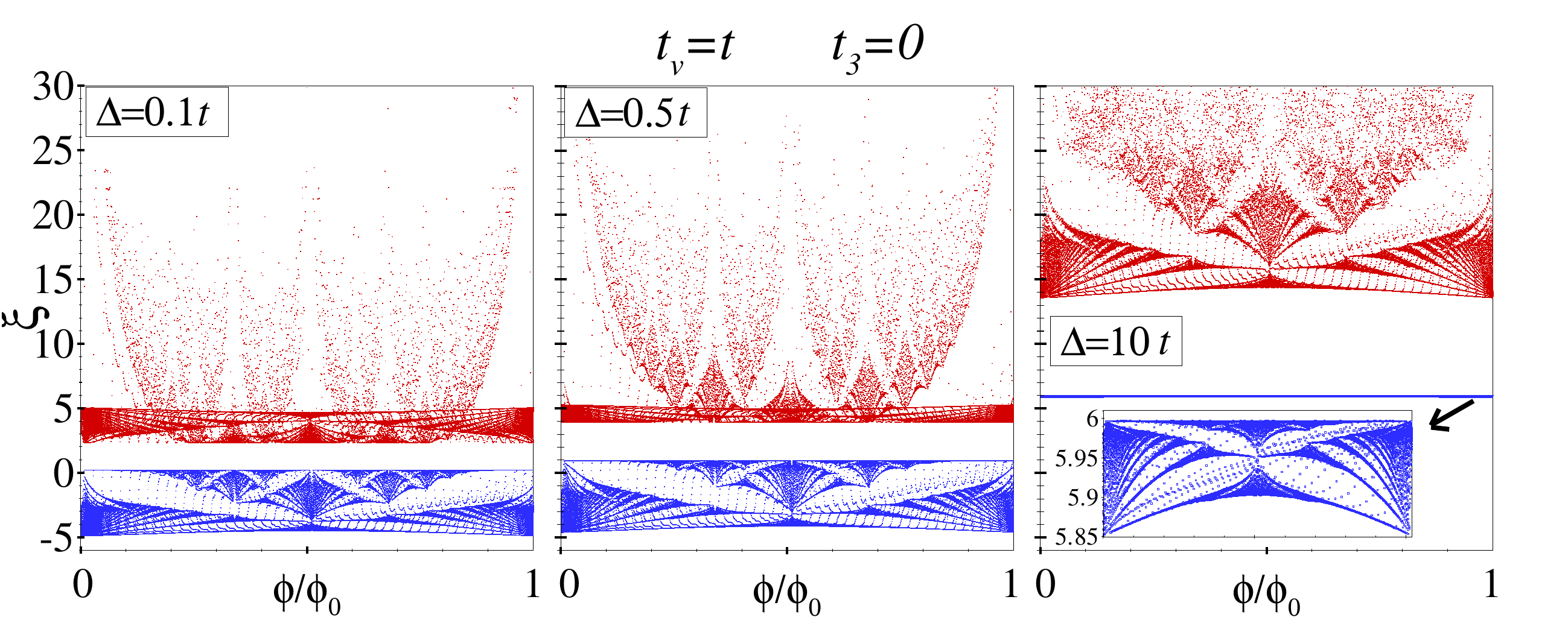}
\caption{Self similar structure of the entanglement spectrum of fermionic BLH
lattice versus reduced magnetic flux, at $t_3=0$, in the presence of on-site
energy difference, for $0\leq p\leq q=51$. A transition from butterfly to a tree-like
picture is shown to be occurred for non-zero $\Delta$.}
\label{ent3plotD2}
\end{figure*}

\subsection{Butterfly entanglement spectrum}\label{sec:HoF-ent}

In order to obtain the entanglement spectrum of the non-interacting free fermions model on BLH lattice
in the presence of a uniform magnetic field we calculate again the single particle correlations.
The non-zero single particle correlations read
\begin{eqnarray}
\no&&\bra{\psi_m}a_{m\kv}^\dg a_{m\kv}\ket{\psi_m},
\bra{\psi_m}b_{m\kv}^\dg b_{m\kv}\ket{\psi_m},\\
&&\bra{\psi_m}a_{m\kv}^\dg b_{m\kv}\ket{\psi_m},
\bra{\psi_m}b_{m\kv}^\dg a_{m\kv}\ket{\psi_m},
\label{non-zero-corr-mag}
\end{eqnarray}
where $a_{m\kv}^\dg$ and $b_{m\kv}^\dg$ ($a_{m\kv}$ and $b_{m\kv}$)
are creation (annihilation) operators of fermions on the lower
layer and $\ket{\psi_m}$ is the ground state of $H_{m\kv}$. Using
these non-zero correlations one can build the entanglement
Hamiltonian of each structural unit cell as
\begin{equation}
{\cal H}_m=\sum\limits_\kv u_{m\kv}^a a_{m\kv}^\dg
a_{m\kv}+u_{m\kv}^b b_{m\kv}^\dg b_{m\kv}+(v_{m\kv} a_{m\kv}^\dg
b_{m\kv}+h.c.), \label{NDEntH-mag}
\end{equation}
where the coefficients $u_{m\kv}^a$, $u_{m\kv}^b$, and $v_{m\kv}$
could be obtained in terms of the nonzero single particle
correlation functions (\ref{non-zero-corr-mag}). By defining
\begin{equation}\label{ent-spect-mag}
\xi_{m\kv}^\pm=\frac
12\left(u_{m\kv}^a-u_{m\kv}^b\pm\sqrt{(u_{m\kv}^a-u_{m\kv}^b)^2+4|v_{m\kv}|^2}\right),
\end{equation}
and using the following unitary transformation:
\begin{eqnarray}
&&\no
a_{m\kv}=\frac{(\xi_{m\kv}^-/v_{m\kv}^*)\c}{\sqrt{1+(\xi_{m\kv}^-/|v_{m\kv}|)^2}}+
\frac{(\xi_{m\kv}^+/v_{m\kv}^*)\d}{\sqrt{1+(\xi_{m\kv}^+/|v_{m\kv}|)^2}},\\
&&b_{m\kv}=\frac{\c}{\sqrt{1+(\xi_{m\kv}^-/|v_{m\kv}|)^2}}+
\frac{\d}{\sqrt{1+(\xi_{m\kv}^+/|v_{m\kv}|)^2}},\label{unitary-trans-mag}
\end{eqnarray}
the Hamiltonian (\ref{NDEntH-mag}) is diagonalized as
\begin{eqnarray}
{\cal H}_m^d=\sum_\kv(\xi_{m\kv}^+\alpha_{m\kv}^\dg\alpha_{m\kv}+\xi_{m\kv}^-\beta_{m\kv}^\dg\beta_{m\kv}).
\end{eqnarray}
Here, $\xi_{m\kv}^\pm$ are the entanglement spectra which are given
by
\begin{equation}
\xi_{m\kv}^\pm=2{\rm arctanh}(2n_{m\kv}^\pm), \label{ent-spect-gen3}
\end{equation}
where
\begin{eqnarray}
\no n_{m\kv}^\pm&=&\frac 12\bigg(\braket{a_{m\kv}^\dg a_{m\kv}}+\braket{b_{m\kv}^\dg b_{m\kv}}\\
\no&\pm&\sqrt{(\braket{a_{m\kv}^\dg a_{m\kv}}-\braket{b_{m\kv}^\dg
b_{m\kv}})^2+4|\braket{a_{m\kv}^\dg b_{m\kv}}|^2}\bigg).
\end{eqnarray}
In the absence of on-site energy difference, at zero skew hopping the ES versus reduced magnetic flux $\phi/\phi_0$,
is perfectly symmetric with respect to $\xi=0$, with a self similar structure like Hofstadter butterfly.
Increasing interlayer vertical hopping causes $\xi^+$ and $\xi^-$ get closer to each other so that at large values of $t_v$ they collapse to $\xi=0$.
The butterfly-like ES at $t_v\cong 2.82 t$ is exactly identical to the MES (compare Fig. \ref{HoFmono} and the right plot in upper panels of Fig. \ref{hofd0}).

In the presence of trigonal warping the symmetry of butterfly-like ES with respect to $\xi=0$ is shown to be broken (see Fig. \ref{hofd0}, the lower panels).
For small skew hoppings, the entanglement levels close to $\xi=0$ start to collapse to $\xi=0$ which causes the separation of the two halves of ES.
At strong skew hopping, a complete collapse to zero entanglement occurs.

In the presence of on-site energy difference, an asymmetry appears in entanglement spectra with respect to $\xi=0$
and an increase of the entanglement is observed.
By increasing $\Delta$ the entanglement levels in the lower half (see Fig. \ref{ent3plotD2}, blue levels)
get closer to each other, whereas the levels in upper half are shown to be spread forming a self similar {\it tree}-like picture.
This butterfly-tree transition is due to the symmetry breaking of single particle distributions on sublattice $A$ and $B$ of the lower layer.

\section{Conclusions and outlook}\label{sec:con}
In this paper we have studied analytically the energy and
entanglement spectrum of a non-interacting free fermions model on
bilayer honeycomb lattice in the presence of trigonal warping on energy spectrum, on-site energy difference and external magnetic field. The fermionic
bilayer honeycomb lattice is somewhat different from the models
considered so far in that the Hamiltonian of layers does not have a
diagonal form. We have proposed a way for determining the
entanglement Hamiltonian of this system by obtaining single particle
correlations of one of both layers on the ground state of the
composite system. Employing the non-zero single particle
correlations we have constructed the entanglement Hamiltonian in
terms of these single particle operators. Diagonalizing this
Hamiltonian we have found out that in the absence of trigonal
warping, at the special value of interlayer vertical coupling
$t_v=2.82 t$ the entanglement spectrum is exactly identical to the
edge state energy spectrum. Trigonal warping breaks down this
correspondence, however, in $\Gamma$-K direction in particular
relevances between hopping parameters the entanglement spectrum is
remarkably the same as monolayer energy spectrum. Moreover similar
to the energy spectrum of the bilayer honeycomb lattice, trigonal
warping also exist on the entanglement spectrum. We have also
studied the effects of on-site energy difference on the entanglement
spectrum of the bilayer and found out that an indirect entanglement
gap is opened on ES by on-site energy difference. The behavior of
this gap is completely different for bilayers with and without
trigonal warping.

In the second part of this paper, we have studied the effects of an
external perpendicular magnetic field on the energy and entanglement
spectrum of the fermionic bilayer honeycomb lattice. Solving Harper's equation and obtaining the non-zero single particle correlations in the ground state of
composite system, we present a non-diagonal entanglement Hamiltonian for Hofstadter bilayer honeycomb lattice. We have shown
that the entanglement spectrum versus magnetic flux is perfectly symmetric and, in correspondence with the energy spectrum, possesses the Hofstadter butterfly
structure. Finally we have shown that in the presence of an on-site
energy difference a transition form butterfly to tree-like picture
occurs on entanglement spectrum.

Bilayer graphene is one of the well-known materials with honeycomb
lattice which has became the focus of numerous theoretical and
experimental works. In bilayer graphene the experimental values \cite{experimentalvaluesgraphene} of hopping parameters are $t=3.16$ eV, $t_v=0.381$ eV, and $t_3=0.38$ eV.
Although they are not in the range where the entanglement spectrum of the bilayer is identical with the
energy spectrum of a single layer graphene, similar to the energy spectrum of bilayer graphene the trigonal warping also seen on the entanglement spectrum of the graphene.

Study of the entanglement properties of fermionic bilayer honeycomb
lattice with other stacking in the presence of in-plane magnetic
field and effects of fermionic interactions are left for the future
works.

{\it Note added}:
During preparing this manuscript we became aware of the paper by Predin, et. al. \cite{Predin15} where the
effects of trigonal warping on the entanglement spectrum of bilayer graphene in the absence of on-site energy difference have been investigated.
The authors have found that although the entanglement spectrum shows qualitative geometric differences
to the energy spectrum of a graphene monolayer, topological quantities such as Berry-phase-type contributions to Chern numbers agree.


\end{document}